# Extending Refusal Testing by Stochastic Refusals for Testing Non-deterministic Systems

Kenza Bouaroudj[1], Ilham Kitouni[1], Hiba Hachichi[1] and Djamel-Eddine Saidouni[1]

1 MISC Laboratory, University Mentouri
Constantine, 25000, Algeria
{ bouaroudj, kitouni, hachichi, saidouni }@misc-umc.org

**Abstract**
Testing is a validation activity used to check the system's correctness with respect to the specification. In this context, test based on refusals is studied in theory and tools are effectively constructed. This paper addresses, a formal testing based on stochastic refusals graphs (SRG) in order to test stochastic system represented by maximality-based labeled stochastic transition systems (MLSTS). First, we propose a framework to generate SRGs from MLSTSs. Second, we present a new technique to generate automatically a canonical tester from stochastic refusal graph and conformance relation $conf_{SRG}$. Finally, implementation is proposed and the application of our approach is shown by an example.

## 1. Introduction

Computer applications become increasingly involved in critical and real-time systems(e.g., automotive, avionic and robotic controllers, mobile phones, communication protocols and multimedia systems). These systems are known by their high complexity. Formal testing allows checking their correctness and helps to ensure their quality.

In order to use a formal testing technique, we need that the systems under study can be expressed in terms of a formal language. During the last two decades, the original formal languages have become more expressive. Thus, a new generation of languages has appeared to allow the explicit representation of non-functional aspects of systems (for example, the probability to perform a task [1][2] or the time consumed by the system to perform task. This time can be fixed [3][4] or defined in probabilistic/stochastic terms[5][6]). The use of stochastic model such as stochastic automata (Network) [8] or stochastic petri net [7]...Etc, allows producing more realistic systems. However, these models are based on interleaving semantics, where the executions of two actions are interpreted by their interleaved executions in time. Following this semantics don't hold when considering action durations with general distributions. We need true concurrency semantic.

In this paper, the system is represented by the "Maximality-based Labeled Stochastic Transition System (MLSTS)" model where actions elapse in time and their durations depend on the probabilistic distributed function. This model is based on maximality semantics [9] and advocates the true concurrency; from this point of view it is well suitable for modeling real time, concurrent and distributed systems.

In this work we are interested in formal testing approaches [10] where the temporal requirements of systems are taken into account. The proposed approach is based on stochastic refusals. Testing based on stochastic refusals allows the comparison between the behavior of the specification and the implementation, if the implementation refuses an action after each timed trace (time is captured by random clock which depends on the probabilistic distributed function), the specification also refuses this action. This theoretical approach is necessary to generate a canonical tester.

In the canonical tester, refusals are found associated with transitions that are led to *"Fail"* location. So, these transitions are labeled by actions which are prohibited by the specification. When the tester is in the location *"Fail"*, this means that the test failed. A test case is a possible path in the tester. Several algorithms exist in literature for the generation of this kind of testers. [11]

### 1.1 Contribution

In this paper we are interested in formal testing approach; we propose a new testing architecture based on stochastic refusals graph (SRG). SRG results from a new definition of refusals. This graph allows us to generate a canonical tester by several transformations of specification graph. Moreover, we investigate the automatic extraction of test cases. The proposed architecture is summarized in Fig.1.

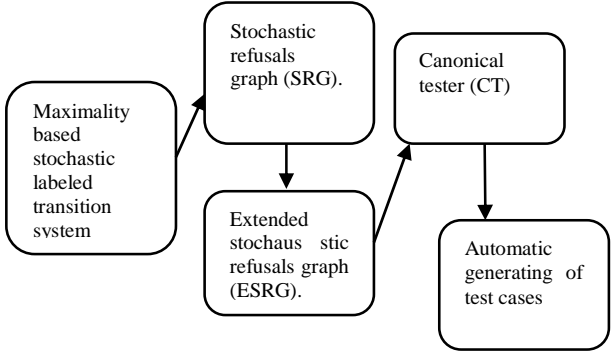

Fig. 1. Test architecture.

First, we calculate sets of refusals; which decorate each state of the specification graph. An important aspect considered at this level is the non-determinism which is captured by permanent refusals.

Temporary refusals are induced by the fact that actions elapse in time. In the MLSTS model the durations of actions are governed by probability distributed functions.

Second, a framework which creates canonical tester from SRG graph and on new conformance relation named "$conf_{SRG}$" is proposed. Finally, an implementation for the proposed frameworks using MATLAB programming language is presented.

### 1.2 Organization

The structure of this paper is as follows: In the next section we start with some informal discussion on maximality semantics and define the maximality based labeled stochastic transition systems model. Section 3 presents conformance relation. We define the new conformance relation $conf_{SRG}$. Section 4 and 5 define the stochastic refusals graph and its extended version. In section 6, the canonical tester is proposed and framework for testing is discussed. In section 7, tool implementation is briefly discussed. Finally, we conclude the paper and discuss some open issues in section 8.

## 2. Maximality Based Labeled Stochastic Transition System (MLSTS)

The semantics associated to the specification model allows the choice of an adequate representation. In the case of the interleaving semantics, the concurrent executions of two actions are interpreted by their interleaved executions in time. Adopting this semantics, every action is supposed to be atomic (structural and temporal) i.e. actions are not divisible and may not elapse in time. These hypotheses make the associated theory simple and the validation tools easy to build.

To escape these hypotheses, true concurrency semantics is defined in the literature [12]. Among these semantics, we can quote the maximality based semantics [13][9][14].

In order to take into account the stochastic aspects, each action of alphabet *Act* are coupled with probability distributed function that governing its duration. An extended action is represented by a pair (*a*, *f*), where *a* is an action type and *f* is the probability distributed function.

The basic idea is to use clock variables to materialize maximum event and to keep track of duration in order to control and observe the passage of time.

Since in our context the action durations are random, clocks are in fact random variables. Transitions are labeled with a clock which represents action's start and appears in the resulting state.

When a clock is reset, it takes a random value whose probability depends on the distributed function of the action duration. As time evolves, clocks countdown synchronously (i.e. all do so at the same rate). When a clock has expired (i.e. has reached the value 0), next transitions are enabled.

The derivation graph obtained with the use of stochastic clocks and by applying the maximality semantics is called maximality based Labeled Stochastic Transition System (MLSTS). [15]

Let *Act* be a finite set of actions and let *DF* be a finite set of probabilistic distributed functions (($\Re\rightarrow[0, 1]$).); *H* is finite set of clocks variables.

**Definition 1(MLSTS):** A maximality based Labeled Stochastic Transition System is a structure $(\Omega, DF, \lambda_S, \mu_S, \psi_S, \xi_S)$ over *Act,* where:

$\Omega = <S, s_0, T>$ is a Transition System with *S*, the countable set of states for the system, $s_0$ is the initial state and *T* is a countable set of transitions specifying the states changes. We define α and β as the same function on the *MLTS*.

- $\lambda_S: T \rightarrow (Act \times DF)$ is a labeling function which associates to each transition a pair *(a, f)* where *a* is an action in *Act* and *f* is probability distributed function in *DF*.

- $\Psi_S: S \rightarrow P(H)$: This function associates to each state a finite set of clocks corresponding to the maximum events.

- $\mu_S: T \rightarrow P(H)$ : This function associates to every transition a finite set of clocks. The expiration of clocks

allows the start of this transition. This set corresponds to the direct causes of it.

- $\xi_S: T \to H$ : This function associates at each transition, a clock that identifies its occurrence and capture duration of actions.

Such that $\psi(s_0)=\emptyset$ and for the all transitions $t$, the functions above have the same properties of those defined in the MLTS model (Subsection 2.1).

Example: consider the MLSTS associated to $E$ is depicted by Fig. 2. After the start of an action $a$ respectively $b$, the state obtained is labeled by $\{x\}$, respectively $\{y\}$. Since actions are independent (parallel), the final state is labeled by $\{x,y\}$.

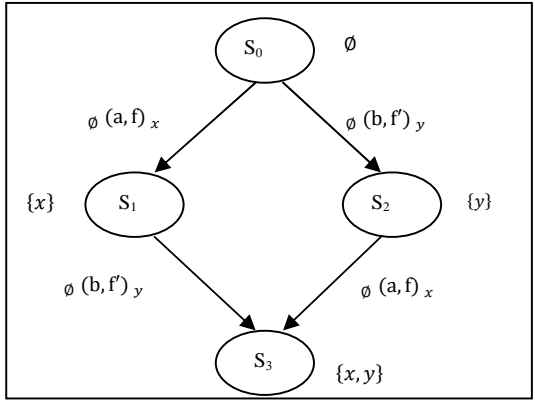

Fig. 2. MLSTS of E.

### 2.1 Non-deterministic MLSTSs

There are several cases to consider when the non-deterministic behavior in MLSTSs is treated. Let $l$ location and $(a,f)$ an extended action. First, if there are two or more transitions from $l$ labeled by $(a,f)$ where all of these transitions have the same set of clocks. Second, if there are two or more transitions from $l$ labeled by $(a,f)$ where all of these transitions have different sets of clocks. Those cases require to be detailed.

Consider two transitions $t, t' \in T: t = (l, M, (a,f), x, l'), t' = (l, N, (a,f), y, l'')$ where M and N are set of clocks.

So,

$$\begin{cases} M = N & (1) \\ M \neq N \begin{cases} M \cap N = \emptyset & (2) \\ M \cap N \neq \emptyset & (3) \end{cases} \end{cases} \quad (1)$$

In fact the first case (Equ.1.1) is an effective non-determinism to consider. The second case (Equ.1.2) is not a non-deterministic case, since there is not choice of transitions. The last case (Equ.1.3) is another non determinism like the first but it has a different significance.

Non-determinism introduces some additional issues: Observability of every possible trace associated with a particular action sequence and the existence of non-equivalent reachable locations.

**Definition 2 (deterministic MLSTS):** maximality based labeled stochastic transition system *Sys* is non-deterministic iff:

$\exists t, t' \in T: \alpha(t) = \alpha(t'), \lambda(t) = \lambda(t')$ and $\mu(t) \cap \mu(t') \neq \emptyset$ (2)

An example is presented in Fig.3.

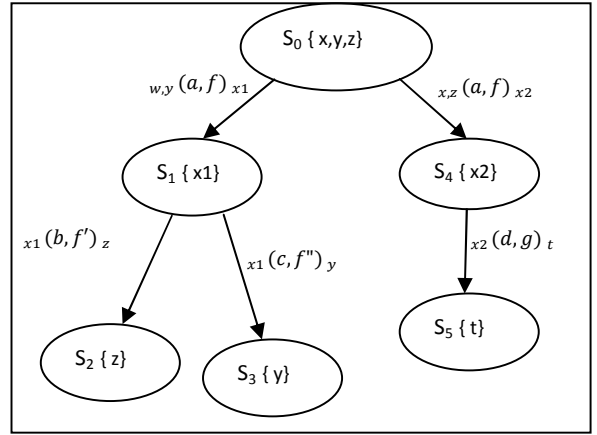

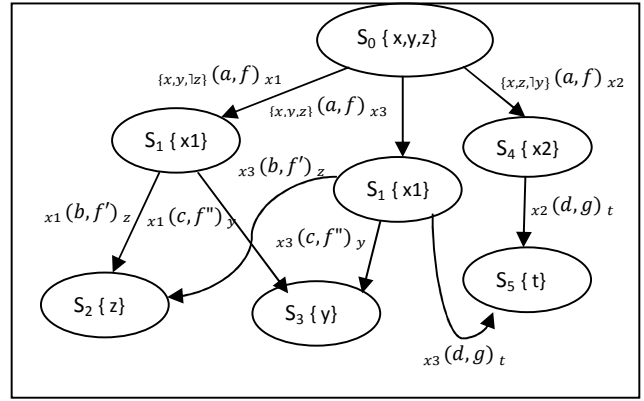

Fig. 3. A Determinized MLSTS.

### 2.2 Determinization of MLSTSs

The determinization application is done in three steps, formally defined in [16].

First step, nondeterministic edges are selected and their clocks sets are determinized two by two.

Second step, a new edge is created and it is labeled by the conjunction of nondeterministic clocks sets.

Third step: outgoing edges of all nondeterministic edges are duplicated and are out coming from the new edge.

## 3. Conformance Relation

In this work we extend the conformance relation *conf* initially defined in [17] and mostly adopted for testing implementations [18].

*Conf relation*: Let *I* and *S* be LTSs, then:

$$I \text{ conf } S \stackrel{def}{=} \forall \sigma \in Traces(S), Ref(I,\sigma) \subseteq Ref(S,\sigma) \quad (3)$$

Intuitively, the *conf* relation holds between an implementation *I* and a specification *S* if, for every trace in the specification, the implementation does not contain unexpected deadlocks. That means that if the implementation refuses an event after such trace, the specification also refuses this event.

To test whether an implementation is conformed to its specification, the notion of canonical tester has been introduced in [8]. A canonical tester for *conf* relation is a labeled transitions system with the same traces as the specification. Moreover, every deadlock of the conformant implementation when it is executed concurrently with the canonical tester has to be a deadlock of the canonical tester [19].

**Definition 2:** Given an LTS $S = (Q, q_0, L, \rightarrow)$, the canonical tester $T(S)$ is defined as follows:

$$\begin{cases} Traces(T(S)) = Traces(S) \\ \forall I, I \text{ conf } S \Leftrightarrow \forall \sigma \in Traces(S), \\ L \in Ref(I\|T(S),\sigma) \Rightarrow L \in Ref(T(S),\sigma) \end{cases}$$
(4)

The operator $\|$ denotes the synchronous (observable events) composition of the LTSs.

Every trace of tester $T(S)$ corresponds to one test case. An execution of $T(S)$ concurrently with an implementation *I* corresponds to a trace of $I\|T(S)$. In order to test conformance, the tester has to be re-executed until all of the traces are passed.

Consequently, before creating a canonical tester, it is necessary to decide when an implementation accord to a specification. For that, we use an implementation relation, and the system is valid when the relation is respected.

All previous approaches are based on interleaving semantics where actions are instantaneous and refusals are permanent.

However, in MLSTS actions have durations and they are governing with probabilities distributed function i.e. the time is stochastic.

The use of stochastic time introduces several technical difficulties in testing. For example the same action may take different amount of time in different run of the system (i.e. the time that system takes to perform actions may vary). So, the conformance relation must be reformulated, we have to take into account actions that system performs and temporal requirements.

One of the possibilities is to consider that any trace of the specification that can be performed by the implementation must have the same duration, that is, an identically distributed random variable. However, in a black-box testing framework, we cannot access to the random variables of the implementation. Consequently, if we consider equivalence of random variables, we need an infinite number of observations from a random variable of the implementation (*with an unknown distribution*) to insure that it has the same distribution as another random variable from the specification (*with a known distribution*). Thus, we have to give more realistic implementation relations based on a finite set of observations.

### 3.1 Conformance Relation conf$_{SRG}$

We define a new conformance relation *conf$_{SPR}$* which is based on the traces and failures of the MLSTS. The extended conformance relation *conf$_{SRG}$* introduce new types of refusals in addition to the classical refusals defined as a set of actions which cannot be permitted from one state. Those refusals are named forbidden actions (*Forb*) to avoid ambiguity. However, the two new kinds of refusals are named: permanent and temporary refusals.

The *Permanent refusals* are generated by the non-determinism in system behavior after the operation of determinization. It is noted $\overline{(a,f')}(X)$

The *temporary refusals* are provoked by actions which elapsed in time. Those refusals are quantified with certain probability, since in MLSTS action duration is governed by probabilistic distributed function. It is noted $\overline{\overline{(a,f')}}(X)$

Given the fact that action duration are represented by a random variable $x_i$, the proposed extended conformance relation is parameterized with definition of $\leq$, this definition means that a distribution in the implementation is conform to a distribution in the specification. Via this parameter, we can allow different choices for $\leq$, as example, we may require the distributions to be equivalent or have the same mean. [20]

**Definition 3:** Let *I* and *S* be MLSTSs

$$Ref_{srg}(g) \stackrel{def}{=} Ref_T(g) \cup Ref_P(g)$$
$$\begin{cases} Ref_T(g) \stackrel{def}{=} temporary\ refusals \\ Ref_P(g) \stackrel{def}{=} permanent\ refusals \end{cases}$$
(5)

So,

$$I\ conf_{srg}\ S \stackrel{def}{=} \forall \sigma \in Traces(S) \begin{cases} (Forb(I,\sigma) \subseteq Forb(S,\sigma))\ and \\ (Ref_{srg}(I,\sigma) \subseteq Ref_{srg}(S,\sigma)\ such\ that\ x_I \leq x_S) \end{cases}$$
(6)

Such as $x_I$ is a random variable representing actions duration of implementation and $x_S$ is random variable representing actions duration of specification.

The use of the ≤ operation between random variables constitutes a general framework that can be instantiated, by giving a specific definition of ≤, when needed [20].

## 4 Stochastic Refusal Graph

In this section we define a Stochastic Refusals Graph (SRG) as structure for testing stochastic non-deterministic systems.

**Definition 4:** A Stochastic Refusals Graph is a deterministic bi-labeled graph structure of MLSTS: $SRG=(\Omega', DF, \lambda_S, \mu_S, \psi_S, \xi_S)$ over an alphabet of extended actions $Q = (Act \times DF)$, where:
  $\Omega'= (G, g_0, \Delta, Ref_{SRG})$ is a Transition System with:
  - G a finite set of localities.
  - $g_0 \in G$ is the initial locality.
  - $\Delta \in (G \times Q \times G)$ is a transition relation. A transition $(g, q, g') \in \Delta$ will also be noted $g \overset{q}{\Rightarrow} g'$.
  - $Ref_{SRG}: G \to P\left(P(\bar{Q} \cup \bar{\bar{Q}})\right)$, is an application that associates for any $g \in G$ a set of refusals where:

$\bar{Q} = \{\bar{q}(X): q \in Q, X \in H \text{ and }\}$ and $\bar{\bar{Q}} = \{\bar{\bar{q}}(X): q \in Q, X \in H\}$

The semantic of the set, $P\left(P(\bar{Q} \cup \bar{\bar{Q}})\right)$ is as follows: Let $g$ be a locality, $g \in G, A \in Ref_{SRG}(g)$

1. $\bar{q} = \overline{(a,f')}(X) \in A$ : is a permanent refusal. It means that an action may be refused permanently to the locality $g$, this refusal is possible but not certain. This certitude will take place after the termination of the actions indexed by (X). See Fig.5.

2. $\bar{\bar{q}} = \overline{\overline{(a,f')}}(X) \in A$ : Means that actions are refused during (X) laps of time. This type of refusals occurs due to the durations of actions. In this case clocks are initialized by a probability distributed function.

As illustration, let consider the maximality based labeled Stochastic transition system $A$ depicted by Fig.4.

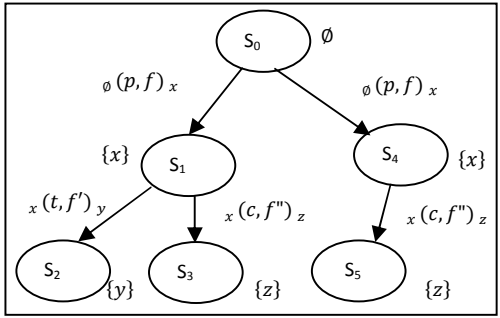

Fig. 4. MLSTS A.

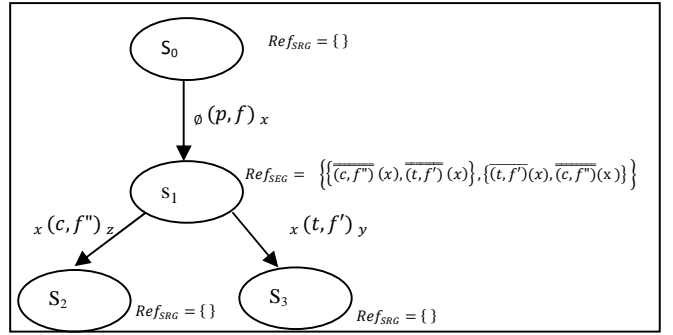

Fig. 5. SRG associated to MLSTS A.

### 4.1 Stochastic Refusals Graphs Generation

In this section we propose a framework to create stochastic refusals graphs from MLSTS specifications.
Let $A = (\Omega, \lambda, \mu, \xi, \psi)$ be an MLSTS, the construction of the SRG associated to $A$ consists to determinize $A$ and decorate all of locations by refusals sets, this requires computing $Ref_{SPR} = \{\cup_{g_i}(\{Pref(g_i) \cup Tref(g_i)\}) : g_i \in G\}$ as follows:
For all location $g$ and for all edges $t$ starting in g:
1. If $t$ is deterministic so:
$$Ref_{SRG}(g) = \begin{cases} Tref(g) = \{\bar{q}(M) | q \in t \text{ and } \mu(t) = M \neq \emptyset\} \\ \cup \, Pref(g) = \emptyset \end{cases}$$
(8)

2. Otherwise: (i.e $t$ is nondeterministic) $E_q(g)$ is a set of non deterministic $t_i$ starting in $g$ and labeling by $q$.
  - Deteriminize $E_q(g)$ and decorate locations by refusals sets as fellows:
    - In step 1 of determinization (subsection2.3), we apply formula Equ.8
    - In step 3 of determinization, let $(g,N,q,y,g')$ be the new edge. We decorate the location $g'$ by :

$Ref_{SRG}(g') = \cup_{t_i} \{\{\cup_k \bar{\bar{q}}_k^i(M_k^i) \cup \{\cup_{j \neq i \wedge k} \bar{q}_k^j(N)\}\}$ :
for all $t_i \in E_q(g)$ and $\beta(t_i) = g_i$ and $t_k^i = (g, M_k^i, q_k^i, x, g_i) \in E(g_i)$  (9)

### 4.2 Stochastic Refusals Graphs Minimization

For minimizing stochastic refusals graphs SRGs, we precede by minimizing refusals sets $Ref_{SRG}(g)$. The minimization procedure of refusals sets eliminates redundant information about refusals at any locality.

**Definition 5:** Let $\Omega' = <G, g_0, \Delta, Ref_{SRG}>$ be a stochastic refusals transition system of SRG, $g$ an element of $G$ and let $A, B$ be elements of $Ref_{SRG}(g)$ :

$A \Subset B$    If    $\forall \bar{a} \in A, \bar{a} \in B$ and $\forall \bar{\bar{a}} \in A$ either $\bar{\bar{a}} \in B$ or $\bar{a} \in B$.       (10)

The minimization of refusals set *A* produces a new set *A'* calculated for any state $g \in G$ is as follows:

1.   $\forall A \in Ref_{SRG}(g) if\ \bar{a} \in A\ and\ \bar{\bar{a}} \in A$     then remove $\bar{\bar{a}}$ from *A*
2.   Minimize $Ref_{SRG}(g)$ with respect to the relation $\Subset$.

In fact, if a set *A* of refusals is an element of $Ref_S(g)$, and both permanent and temporary refusals on action *a* are in *A*. It means that a system may be in a state when action *a* is definitely refused or temporary refused. Then, no way permits to ensure that action *a* will be offered after a laps of time. Which justifies the remove of temporary refusals of action *a* in the set *A*. In the second step, if $A \Subset B$ so *A* is removed. In fact, the refusals in *B* contain the refusals in *A*. The stochastic refusals graph SRG is said minimal if the refusals set $Ref_S$ remains unchangeable by the application of Step1 and Step2 for any locality $g \in G$.

## 5. Extended Stochastic Refusals Graph

Extended refusal set is defined as the extension of location refusals sets $Ref_{SRG}$ by forbidden actions. Forbidden actions are actions which aren't permitted from one location.

**Definition 8:** Let $SRG(S) = (\Omega', DF, \lambda_S, \mu_S, \psi_S, \xi_S)$, the extended transition system of $\Omega'$ is $\Omega'_! = (G, g_0, \Delta, Ref_!)$ defined as the extension of refusals sets of state $Ref_{srg}$ by forbidden actions: $Ref_!(g) = (Ref_{srg}(g) \cup Forb(g))$
The $SRG_!(S) = (\Omega'_!, DF, \lambda_S, \mu_S, \psi_S, \xi_S)$ is named extended stochastic refusals graph.

## 6. Canonical Tester

A canonical tester with respect to *conf*$_{SRG}$ is able to detect every implementation that disagrees with a specification, thus if the implementation refuses an action after each timed trace, the specification also refuses this action. That means, *I* and *S* have the same refusals sets and the same timed traces. This theoretical approach is necessary to generate a canonical tester. In the proposed canonical tester, three verdicts $\{pass, incon, fail\}$ are used. At every step of the test computation if a locality is reachable so it is decorated by *pass* verdict. The inconclusive verdict *incon* is produced by the non-determinism present in the system, and captured by permanent refusals set. *Fail* is a new locality introduced to canalize transitions labeled by actions which are not permitted. Two cases of actions are not allowed: first, when an action is in the forbidden set of state. The second case, when an action is offered without respecting the set of clocks that it depends to. This action is in temporary refusals set.

A framework for creating the canonical tester of MLSTS follows these steps:
- First, we generate SRG from MLSTS as define in the previous section
- Then, we generate the extended stochastic refusals graph $SRG_!$.
- Finally, the canonical tester is constructed over $SRG_!$. All traces of the canonical tester are considered as test cases.

The canonical tester M" associated to SRG of Fig. 5 is depicted by Fig. 6.

Fig.7 represents an example of a test case with inconclusive verdict.

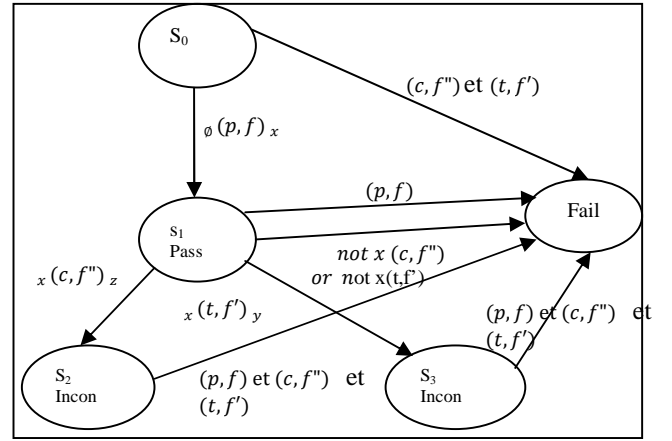

Fig. 6. Canonical tester associated to SRG.

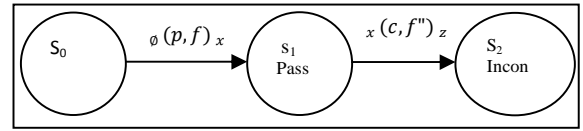

Fig. 7. Test case associated to the canonical tester M".

.

## 7. Implementation

The proposed approach was implemented in MATLAB language. The choice of this language is motivated by the easy and efficient manipulation of vectors and matrices which are the effective representation of SRG. A lot of resourceful toolboxes exist particularly a TORSCHE Scheduling toolbox [21] which contains predefined functions, facilitates the manipulation of interesting data structures also it proposes a useful graphical editor.

To explain how the tool works, let as propose the example of ATM system. This machine allows withdrawing money from account. Its behavior is as follow. Customer has to insert card in machine. After he has to type a code, if code is correct the machine delivers money and card. If the code is wrong, the machine can keep card or reject it.

This machine is depicted by the Fig. 8. Fig.9 presents the same example edited by the graph editor.

We use the implemented tool to generate canonical tester depicted by Fig. 11 over the extended stochastic refusals graph, Fig.10.

In Fig.10, we notice that every locality is decorated. We take the example of the location S4.the location is decorated as follow:

'S4','y1','[(keepcart,g4)`(y1)`,(rejectcart,g2)(y1),][(rejectcart,,g2)`(y1)`,(keepcart,,g4)(y1),]','(incart,f),(valide,f1),(codenotok,g),(codeok,g1),(outcart,g5),(take,g6)'

Where 'S4' is the name of localitie, 'y1' is the set of clocks, '[(keepcart,g4)`(y1)`,(rejectcart,g2)(y1),][(rejectcart,g2)`(y1)`,(keepcart,g4)(y1),]' is $Ref_{srg}$ and ',(incart,f),(valide,f1),(codenotok,g),(codeok,g1),(outcart,g5),(take,g6)' is $Forb$ set.

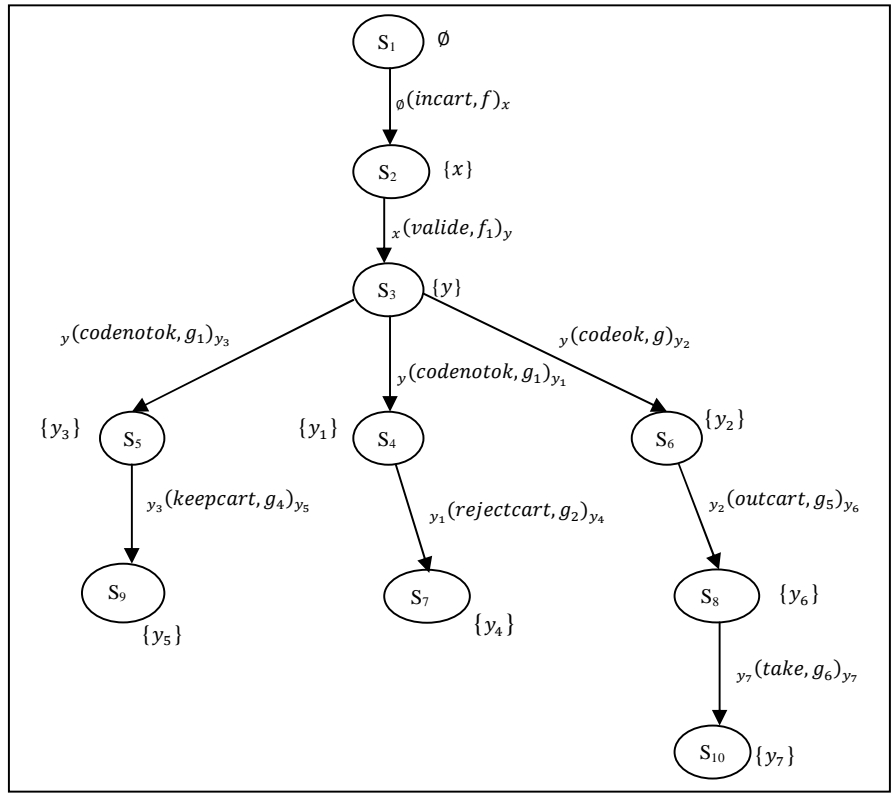

Fig. 8. MLSTS of ATM example.

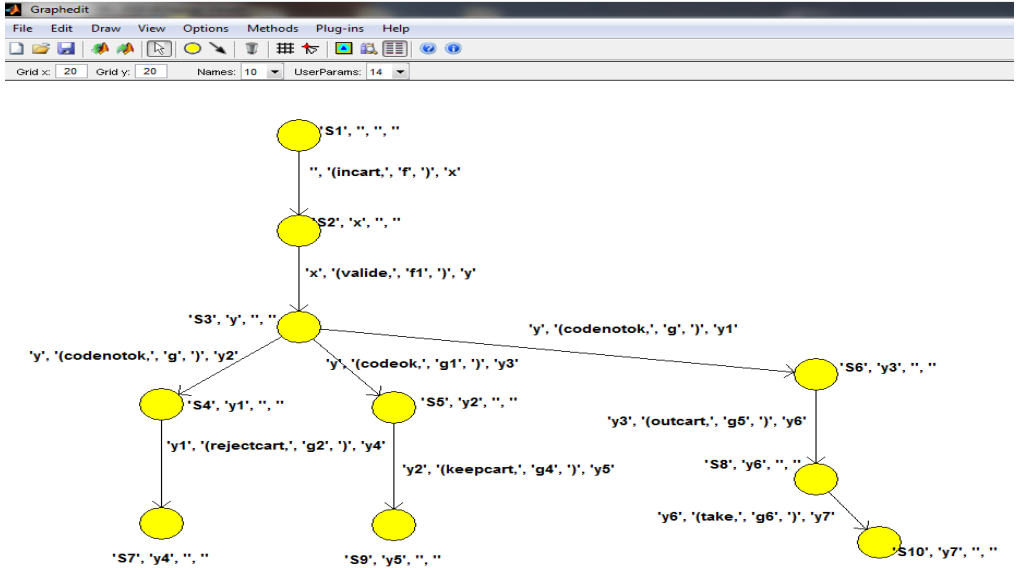

Fig. 9. MLSTS edited by the graph editor.

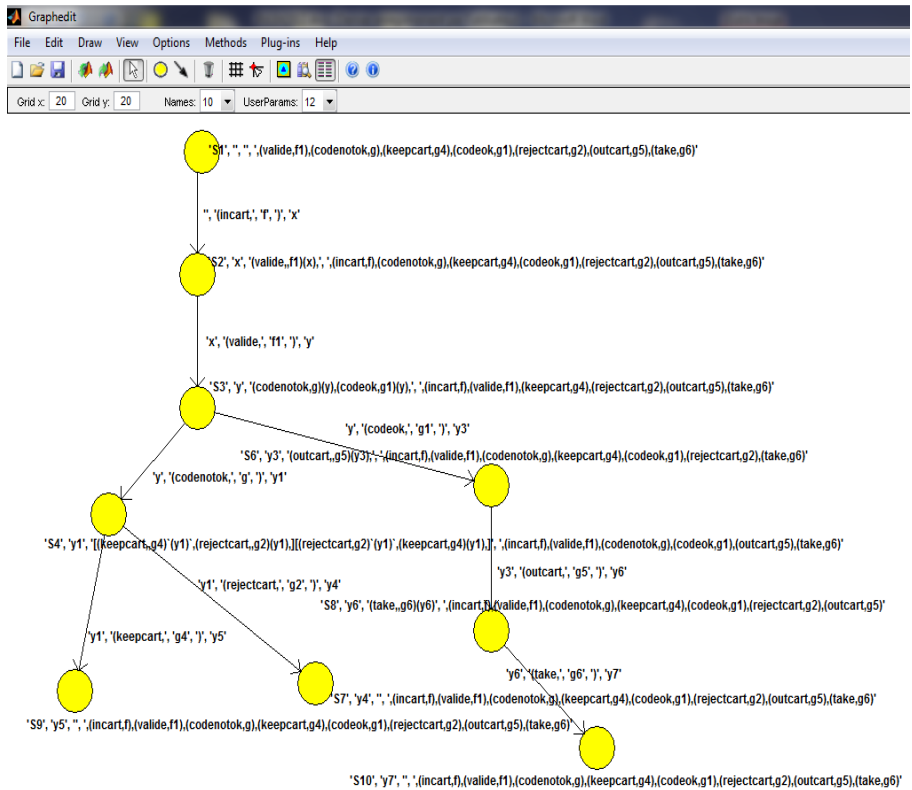

Fig. 10. Extended stochastic refusals graph

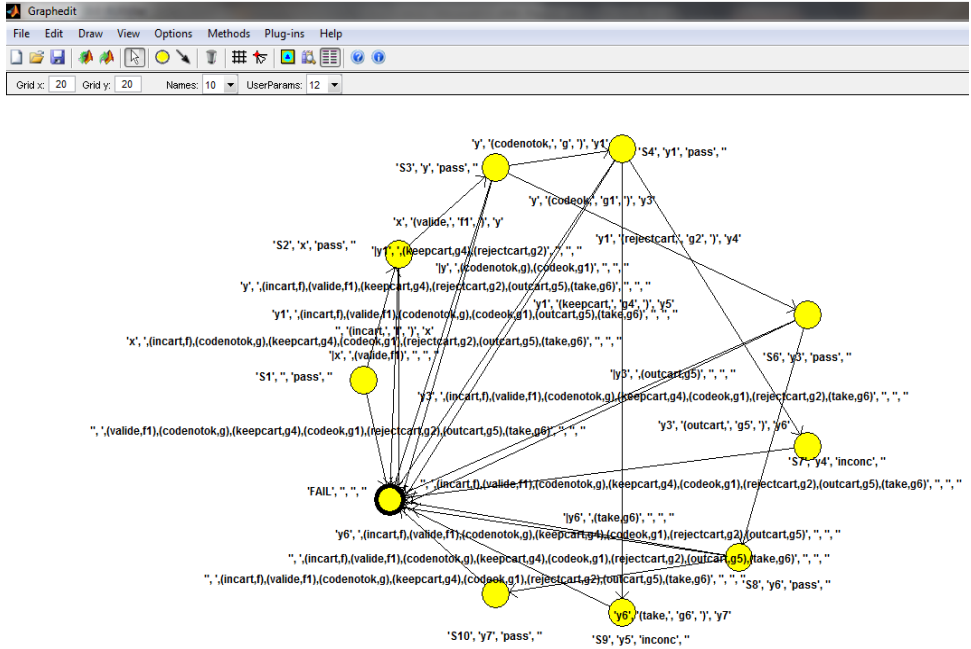

Fig. 11. Canonical tester, CanTes.

In Fig.11 we notice that every location is decorated with verdicts (pass, incon) and new locality is created which is decorated with fail verdicts. The new locality *Fail* is introduced to canalize transitions labeled by actions which are not permitted such as those on  *Forb* set. The framework used to generate this tester is presented in section

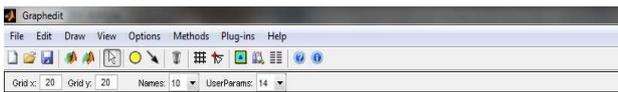
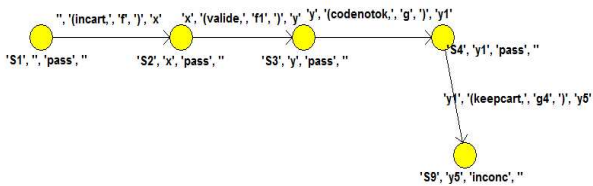

Fig. 12.  Test case.

Fig. 12 is test cases. Test case is path of tester. It is generated randomly

## 8. Conclusion

In this paper, , we have proposed an approach for testing s stochastic systems modeled in Maximality based Labeled Stochastic Transition System (MLSTS). First, we have proposed a stochastic refusals graph for MLSTS specifications. In this variant of refusals graphs, temporary refusals are quantified with a certain probability, because the duration of actions are represented with a probability distributed function. Next we propose framework to generate a canonical tester with automatic extraction of test cases. Finally, an implementation is proposed using MATLAB programming language.

As perspectives, we plan to complete this work by strategy for choosing which of test cases are sufficient for insuring some completeness guarantees. A related problem is how to measure the "goodness" of a set of test cases and how to select test suites with some good coverage measure.

# References

[1] Nicollin, X., Sifakis J.: An overview and synthesis on timed process algebras. In 3rd Int. Conf. on Computer Aided Verification, CAV'91, LNCS 575, pages 376-398. Springer, (1991).

[2] Hennessy, E., Regan, T.: process algebra for timed systems. Information and Computation, 117(2):221-239, (1995)

[3] Lopez, N., Núñez M.: A testing theory for generally distributed stochastic processes. In 12th Int. Conf. on Concurrency Theory, CONCUR'01, LNCS 2154, pages 321-335. Springer, (2001).

[4] Bravetti, M., Gorrieri R.: The theory of interactive generalized semi-Markov processes. Theoretical Computer Science, 282(1):5-32, (2002).

[5] Cazorla, D., Cuartero, F., Valero, V., Pelayo, F.L., Pardo, and J.J.: Algebraic theory of probabilistic and non-deterministic processes. Journal of Logic and Algebraic Programming, 55(1-2):57-103, (2003).

[6] Nuñez, M.: Algebraic theory of probabilistic processes. Journal of Logic and Algebraic Programming, 56(1-2):117-177, (2003).

[7] Bause, P., Kritzinger S: Stochastic Petri Nets -An Introduction to the Theory-, 2nd Edition, Vieweg Verlag, Germany, ISBN: 3-528-15535-3,2002.

[8] D'Argenio, P. R., Katoen, J.-P: A theory of stochastic systems. Part II: Process algebra, In. Information and Computation 203, Elsevier , pp. 39–74 Inc, 2005.

[9] Saïdouni, D.E., Benamira, A., Belala, N., Arfi, F.: FOCOVE: Formal concurrency verification environment for complex systems. In Mediterranean Conference on Intelligent Systems and Application (CISA'2008). Annaba, Algeria, June 29-30th and July 1st 2008.

[10] Hessel, A. , Larsen, K . G. , Mikucionis, M. , Nielsen, B. , Pettersson, P., Skou, A. : Testing Real-Time Systems Using UPPAAL. Formal Methods and Testing,pp 77-117 (2008

[11] J. Tretmans: Testing Concurrent Systems: A Formal Approach. Jos C.M. Baeten, Sjouke Mauw (Eds.): CONCUR'99, LNCS 1664, pp. 46_65, 1999. Springer-Verlag.

[12] Langerk, R.: "Bundle Event Structures: A non-interleaving Semantics for LOTOS " in Diez M. and Groz R (EDS), in proceedings of FORTE 92 , pp. 331-346 North Holland, (1993).

[13] Saïdouni, D. E., Belala, N., Bouneb, M.: Aggregation of transitions in marking graph generation based on maximality semantics for Petri nets. In Proceedings of the 2nd Workshop on Verification and Evaluation of Computer and Communication Systems (VECoS'2008), University of Leeds, UK. BCS, July, 2-3rd (2008).

[14] Saïdouni, D. E., Belala, N.: Using maximality-based labeled transition system model for concurrency logic verification. The International Arab Journal of Information Technology (IAJIT), 2(3):199—205, ISSN: 1683-3198, July (2005).

[15] Arous, M., Saidouni, D. E., Ilié, J. M.: Maximality Semantics based Stochastic Process Algebra for Performance Evaluation.1st IEEE International Conference on Communications, Computing and Control Applications (CCCA'11) March 3-5, at Hammamet, Tunisia. IEEE Catalog Number: CFP1154M-ART, ISBN: 978-1-4244-9796-6 (2011).

[16] Saidouni, D.E., kitouni, I., Bouarroudj, k., Hachichi, h. : Extending Refusal Testing by Temporary and Permanent Refusals for Non-deterministic Systems. Reschrch repport 12003, Misc laboratory, mentouri university Constantine, 25000, algeria

[17] E. Brinksma. A theory for the derivation of tests. In S. Aggarwal and K. Sabnani, editors, Proceedings of the 8th IFIP Symposium on Protocol Specification, Testing and Verification (PSTV 1988). North-Holland, 1989

[18] A. Cavalcanti and M-C. Gaudel. : A note on traces refinement and the conf relation in the Unifying Theories of Programming. In Andrew Butterfield, editor, Unifying Theories of Programming, Second International Symposium, UTP 2008, Trinity College, Dublin, Ireland, September 8-10, (2008).

[19] M. Broy, B. Jonsson, J-P. Katoen, M. Leucker, A. Pretschner (Eds.): Model-Based Testing of Reactive Systems, LNCS 3472, Chap. 6, pp. 151-171, Springer-Verlag Berlin Heidelberg 2005.

[20] Robert, M., Hierons, Mercedes G. Merayo, Manuel Núñez: Testing from a stochastic timed system with a fault model. J. Log. Algebr. Program. 78(2): 98-115 (2009)

[21] Šůcha, P., Kutil, M., Sojka, M., Hanzálek, Z.: TORSCHE Scheduling Toolbox for Matlab. In IEEE International Symposium on Computer-Aided Control Systems Design. Munich, Germany: (2006).

**Kenza Bouaroudj** received her master's degree in computing sciences from University of Mentouri Constantine, Algeria in 2010. Currently, she is a PhD student at CFSC research group of MISC laboratory, Mentouri University of Constantine, Algeria. Her research interests are system validation and testing real-time stochastic systems.

**Ilham Kitouni** obtained her BEng degree from University of Mentouri Constantine, Algeria, in 1992, after 15 years in different Algerian company as head of department of Computer Sciences, she recovers CFSC research group of MISC laboratory, Mentouri University of Constantine, Algeria. From October 2009, she prepares a PhD thesis. Her research domain is formal models for real-time systems specification and validation.

**Hiba Hachichi** received her master's degree in computing sciences from University of Mentouri Constantine, Algeria in 2009. Currently, she is a PhD student at CFSC research group of MISC laboratory, Mentouri University of Constantine, Algeria. Her research interests are graph transformation and formal methods for verifying and testing real time systems.

**Djamel-eddine Saidouni** received his PHD degree in theoretical computer science from the university Paul Sabatier of Toulouse, France in 1996. Actually he is a professor at the department of computer science, Mentouri University of Constantine, Algeria. Also, he is the head of the CFSC research group of MISC laboratory. His main research domain interests formal models for specifying and verifying critical systems, real time systems, true concurrency models and state space explosion problem.